\documentclass[useAMS,usenatbib,usegraphicx]{mn2e}
\usepackage{longtable}
\title[Circumstellar discs in the young $\sigma$\,Orionis cluster]{Circumstellar discs in the young $\sigma$\,Orionis cluster}
\author[Oliveira, Jeffries, van Loon \& Rushton]{J.M. Oliveira\thanks{E-mail:
joana@astro.keele.ac.uk}, R.D. Jeffries, J.Th. van Loon and M.T. Rushton\\
School of Physical and Geographical Sciences, Lennard-Jones Laboratories, Keele 
University, Keele, Staffordshire ST5 5BG, UK}
\begin{document}

\date{}

\pagerange{\pageref{firstpage}--\pageref{lastpage}} \pubyear{2005}

\maketitle

\label{firstpage}

\begin{abstract}
We present new K- and L$'$-band imaging observations for members of the young
(3$-$5\,Myr) $\sigma$\,Orionis cluster, obtained with UIST at UKIRT. We 
determine $(K-L')$ colour excesses with respect to the photospheres, finding 
evidence for warm circumstellar dust around 27 out of 83 cluster members that 
have masses between 0.04\,M$_{\odot}$ and 1.0\,M$_{\odot}$. This indicates a 
circumstellar disc frequency of at least (33\,$\pm$\,6)\% for this cluster, 
consistent with previous determinations from smaller samples 
\citep*{oliveira04b} and also consistent with the 3\,Myr disc half-life 
suggested by \citet*{haisch01b}. There is marginal evidence that the disc 
frequency declines towards lower masses, but the data are also consistent with 
no mass-dependence at all. There is no evidence for spatial segregation of 
objects with and without circumstellar discs.
\end{abstract}

\begin{keywords}
circumstellar matter -- infrared: stars -- stars: pre-main-sequence -- stars:
low-mass, brown dwarfs -- open clusters and associations: individual 
($\sigma$ Orionis)
\end{keywords}

\section{Introduction}

Circumstellar discs play an important role in the formation and early evolution
of low-mass stars. There is ample empirical evidence that the majority of stars
are born with circumstellar discs \citep*{haisch00,lada00}. It has also become 
clear that dust discs are removed relatively quickly, likely within 10\,Myr 
\citep*{haisch01b}. As the building blocks for planetary systems, it is 
imperative that we understand disc processes. In particular the timescale for 
disc dissipation is crucial in determining whether planets form and on what 
timescales \citep{brandner00} and might control exoplanet growth and migration 
\citep{lecar03}. It is also not well understood how the star formation
environment might condition disc evolution and consequently planet formation.

Recently, a huge observational effort has been channelled into determining the 
disc frequency in star forming regions, in order to empirically constrain the 
disc destruction timescale. Circumstellar dust discs are cooler than the stellar
photospheres, therefore they irradiate mainly at infrared (IR) wavelengths. In 
particular, a $(K-L)$ colour excess is a robust disc indicator \citep{wood02}, 
at least for the first few Myr; for older associations the inner disc becomes 
optically thin and dust excesses are only detected at longer wavelengths 
\citep{uchida04,megeath05}. Using $(K-L)$ excess as a disc indicator, 
\citet{haisch01b} determine that within the first 3\,Myr approximately half of
the dust discs disappear and that at 6\,Myr most low-mass stars have lost their
discs. This timescale relies on few clusters with rather uncertain ages; 
observations of more clusters in the crucial 3$-$5\,Myr age range are essential.

The $\sigma$\,Orionis association was first identified in {\em ROSAT} images as
a concentration of bright X-ray sources around the massive multiple system 
\citep{wolk96}. Photometric surveys \citep*{bejar01,sherry04} identified a 
score of pre-main-sequence (PMS) candidates well into the brown dwarf regime and
\citet{osorio00} discovered several objects with masses below the deuterium 
burning limit \citep[for ongoing surveys on very-low-mass cluster candidates see][]{caballero05}. 
Optical spectroscopy has been used to confirm membership for many
cluster candidates \citep{osorio02,barrado03,kenyon05}. \citet{scholz04} 
investigate variability and rotation in very low-mass cluster members while 
\citet*{franciosini05} analysed the XMM-{\em Newton} X-ray properties of PMS 
stars within the central 30\,arcmin area. The age of the cluster is 3$-$5\,Myr 
\citep*[e.g.,][]{osorio02,oliveira04b} for an Hipparcos distance (to the central 
star) of 352\,pc. The reddening towards the cluster namesake is low, 
$E(B-V) = 0.05$ \citep[e.g.,][]{brown94}.

We have been investigating the IR properties of the $\sigma$\,Ori cluster 
members. \citet{loon03} and \citet{oliveira04a} use mid-IR imaging and
spectroscopy to confirm several objects as classical T\,Tauri stars (CTTS) and 
therefore likely members of the $\sigma$\,Ori cluster. \citet{oliveira04b} use 
$(K-L')$ excesses for a sample of 24 cluster members to constrain the cluster 
disc frequency; they found that $(46 \pm 14)$\% of cluster members retain their 
discs. We have since expanded that thermal-IR survey; we here present new 
imaging observations of an additional 59 cluster members in the K- and 
L$'$-bands. We compute the disc frequency for the complete sample of 83 cluster 
members and investigate the mass dependence of disc dissipation across a wider 
mass range.

\section{Sample of cluster members}
\label{target_sample}

Targets were selected from several samples from the literature. Objects from 
\citet{osorio02}, \citet{barrado03}, \citet{muzerolle03} and \citet{kenyon05} 
have been confirmed as cluster members using a combination of spectroscopic 
indicators: the presence of the Li\,{\sc i} 6708\,\AA\ line, a weak Na\,{\sc i} 
IR doublet and radial velocity measurements. We did not choose members 
identified on the basis of strong H$\alpha$\ emission in order not to bias the 
sample towards objects with accretion discs. To complement the sample at the 
higher mass end we selected high photometric probability ($>$\,70\,\%) members 
from the optical and near-IR sample of \citet{sherry04}. There are also 2 
targets from the photometric sample of \citet{bejar01}. 

\begin{figure}
\includegraphics[height=9.5cm]{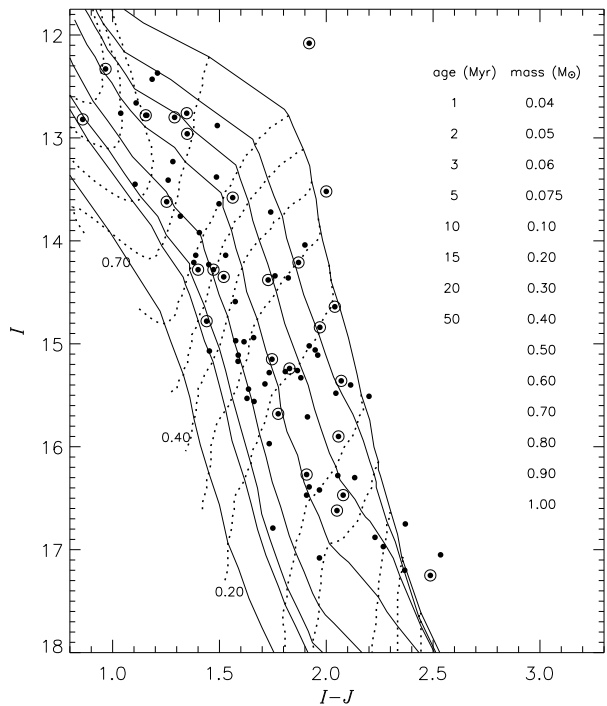}
\caption{$I/(I-J)$ colour-magnitude diagram of the observed $\sigma$\,Ori 
cluster members, from \citet{oliveira04b} and new observations; double-circle
objects exhibit a 2-$\sigma$ $(K-L')$ excess (see Sect.\,\ref{excess}). 
Isochrones and evolutionary tracks are from \citet{baraffe98} and the Hipparcos
distance of 352\,pc is used. No extinction correction is performed (see text).}
\label{cmd}
\end{figure}

Fig.\,\ref{cmd} shows the $I/(I-J)$ colour-magnitude diagram of the observed 
sample as well as the sample from \citet{oliveira04b}, together with isochrones 
and evolutionary tracks from \citet{baraffe98}, adopting the Hipparcos distance 
of 352\,pc. \citet{oliveira04b} investigated whether differential reddening 
should be taken into account; they found that the only objects to exhibit 
significant reddening also display an IR excess. We could use the reddening 
determination towards $\sigma$\,Ori itself, but it is rather small \citep[$E(B-V) = 0.05$,][]{brown94}
in particular at IR wavelengths; therefore we follow \citet{oliveira04b} and do 
not de-redden targets' magnitudes and colours. The masses of the cluster members
are then in the range 0.04$-$1.0\,M$_{\odot}$, with 6 objects below the brown 
dwarf boundary. The median age of the sample is 4.2\,Myr, consistent with 
previous determinations, e.g., 4.2$^{+2.7}_{-1.5}$\,Myr from 
\citet{oliveira02}. If a larger distance of 440\,pc is adopted, the isochronal 
age of the cluster decreases to 2.5\,Myr \citep{sherry04} --- and the computed 
PMS masses are only slightly increased. In any of these studies the presence of
hitherto unidentified binaries can cause the age determination to be biased 
towards younger ages. A small average reddening would also cause the cluster to 
appear younger, but in the case of the $\sigma$\,Ori cluster this effect is 
negligible. From the absence of lithium  depletion in the cluster members, 
\citet{osorio02} estimate an upper limit to the cluster age of 8\,Myr. Cluster 
members also appear to present a large age spread (see discussions in 
\citealt{oliveira04b} and \citealt{burningham05}). The effect of different age 
estimates is taken into account later in this discussion.

\section{Observations: K and L$'$-band imaging}

The K- and L$'$-band imaging was performed with the imager spectrometer UIST at
the United Kingdom Infrared Telescope (UKIRT) over semesters 2003B and 2005A. 
These observations closely follow the strategy described in \citet{oliveira04b}
therefore we just summarise the few major points. 

K$_{\rm s}$-band magnitudes from 2MASS (Two Micron All Sky Survey, 
\citealt{cutri03}) were used to set the exposure times in this band, while 
evolutionary models from \citet{baraffe98} provided estimates of the L-band 
photospheric magnitudes of the targets, in the absence of circumstellar
emission. Observations were performed under UKIRT flexible scheduling rules, so 
that conditions were optimal (dry conditions and seeing better than 
0.6\,arcsec). Total exposure times in the L$'$-band range from 
$\sim$2$-$60\,min, while K-band exposures lasted 1\,min. To minimise the 
effects of variability, observations in the two filters were consecutive. Images
were reduced and combined using ORAC-DR (the UKIRT data reduction and high 
level instrument control software) and aperture photometry was performed 
with GAIA (Graphical Astronomy and Image Analysis Tool), both Starlink packages.
Calibration onto the MKO-NIR (Mauna Kea Observatory Near-Infrared) system 
\citep*{tokunaga02} was achieved using photometric standards \citep[][ and also the UKIRT\footnotemark web page]{leggett03}:
SAO\,112626 ($L'=8.56 \pm 0.01$\,mag) and FS\,11 ($K=11.241 \pm 0.008$\,mag), 
where uncertainties are derived from multiple measurements. 

\footnotetext{http://www.jach.hawaii.edu/UKIRT/astronomy/}

\begin{table*}
\centering
\begin{minipage}{180mm}
\caption{New photometry of $\sigma$\,Ori cluster members. Column 1 is the target 
number, columns 2 and 3 are the targets' 2MASS positions, column 4 is the
$I_{\rm c}$ magnitude, columns 5$-$10 are the 2MASS $J,H,K_{\rm s}$ magnitudes 
with uncertainties, columns 11$-$14 are the new MKO-IR $K$ and $L'$ magnitudes 
with uncertainties and column 15 gives the target identification. Objects with
a label S or K, followed by an ID number, were identified respectively by
\citet{sherry04} or \citet{kenyon05}. Object IDs like 4771$-$1097 and 
r053833$-$0236 are X-ray sources from \citet{wolk96} --- see also 
\citet{osorio02}.}
\label{obs_table}
\begin{tabular}{@{\hspace{-0mm}}r@{\hspace{3mm}}c@{\hspace{2mm}}c@{\hspace{2mm}}l@{\hspace{2mm}}r@{\hspace{2mm}}l@{\hspace{2mm}}r@{\hspace{2mm}}l@{\hspace{2mm}}r@{\hspace{2mm}}l@{\hspace{2mm}}r@{\hspace{2mm}}l@{\hspace{2mm}}r@{\hspace{2mm}}l@{\hspace{2mm}}l}
\hline
  &    RA     &     DEC  &  \multicolumn{1}{c}{$I_{\rm c}$} & \multicolumn{2}{c}{$J$}   &   \multicolumn{2}{c}{$H$}  &\multicolumn{2}{c}{K$_{\rm s}$}&\multicolumn{2}{c}{$K$} & \multicolumn{2}{c}{$L'$}&\multicolumn{1}{l}{Identification} \\
  & ($^{h\,\,m\,\,s}$)&($^{d\,\,m\,\,s}$) & (mag) & \multicolumn{2}{c}{(mag)}&\multicolumn{2}{c}{(mag)}  & \multicolumn{2}{c}{(mag)} & \multicolumn{2}{c}{(mag)}&\multicolumn{2}{c}{(mag)} \\
\hline
 1&05 39 39.38&$-$2 17 04.5&12.96&11.611&0.023&10.714&0.023&10.172&0.024&10.207&0.009& 9.193&0.040      &S 33  		  \\
 2&05 38 35.87&$-$2 30 43.3&12.43$^{*}$&11.245&0.026&10.598&0.023&10.424&0.024&10.429&0.009&10.257&0.029&4771$-$1097	  \\ 
 3&05 38 44.23&$-$2 40 19.7&12.33$^{*}$&11.363&0.026&10.688&0.024&10.439&0.024&10.382&0.009& 9.651&0.027&4771$-$1051	  \\
 4&05 38 39.82&$-$2 56 46.2&12.76$^{*}$&11.413&0.027&10.744&0.023&10.439&0.021&10.452&0.009& 9.664&0.026&S 145            \\
 5&05 38 47.92&$-$2 37 19.2&13.58&12.018&0.042&11.239&0.046&10.776&0.037&10.945&0.010&10.080&0.026      &S 102  		     \\
 6&05 38 53.07&$-$2 38 53.6&12.78$^{*}$&11.625&0.026&11.034&0.026&10.828&0.025&10.780&0.010&10.455&0.022&S 166  		     \\
 7&05 37 24.27&$-$2 19 07.6&12.76&11.722&0.024&11.039&0.024&10.844&0.023&10.846&0.009&10.707&0.070      &S 99  		     \\
 8&05 37 51.61&$-$2 35 25.7&13.38&11.894&0.026&11.172&0.023&10.977&0.022&11.009&0.010&10.796&0.054      &S 125  		     \\
 9&05 39 08.53&$-$2 51 46.6&13.23&11.948&0.024&11.201&0.023&11.028&0.024&10.982&0.010&10.812&0.020      &S 61  		     \\
10&05 38 08.27&$-$2 35 56.3&13.64&12.142&0.026&11.376&0.023&11.047&0.019&11.265&0.010&10.801&0.024      &S 41  		     \\
11&05 38 34.06&$-$2 36 37.5&13.72&11.980&0.027&11.330&0.024&11.077&0.027&10.988&0.001&10.612&0.025      &r053833$-$0236    \\ 
12&05 39 33.79&$-$2 20 39.9&13.62&12.367&0.026&11.598&0.023&11.429&0.023&11.495&0.010&11.235&0.015      &S 138  		     \\
13&05 36 29.09&$-$2 35 48.3&13.76$^{*}$&12.443&0.024&11.689&0.031&11.495&0.023&11.438&0.010&11.429      &0.064&S 143  		     \\
14&05 38 37.94&$-$2 05 52.4&13.45&12.345&0.027&11.723&0.024&11.502&0.023&11.500&0.010&11.280&0.043      &S 168  		     \\
15&05 38 58.55&$-$2 15 27.8&13.92&12.514&0.027&11.790&0.024&11.551&0.021&11.570&0.010&11.337&0.096      &S 46 		     \\
16&05 38 36.69&$-$2 44 13.7&14.36&12.538&0.027&11.891&0.026&11.623&0.028&11.583&0.011&11.264&0.054      &S 16 		     \\
17&05 38 20.50&$-$2 34 09.0&14.38&12.652&0.026&11.918&0.023&11.648&0.019&11.810&0.010&11.044&0.022      &r053820$-$0234   \\ 
18&05 39 02.77&$-$2 29 55.8&14.14&12.611&0.028&12.001&0.024&11.694&0.023&11.657&0.011&11.452&0.088      &S 28  		     \\
19&05 38 18.86&$-$2 51 38.8&14.28&12.808&0.023&12.039&0.023&11.733&0.021&11.750&0.011&11.137&0.112      &S 39  		     \\
20&05 38 23.65&$-$3 01 33.2&14.14&12.751&0.023&12.147&0.026&11.908&0.024&11.884&0.011&11.823&0.050      &S 38  		     \\
21&05 37 54.05&$-$2 44 40.7&14.59&13.016&0.026&12.339&0.026&12.104&0.022&12.058&0.012&12.134&0.200      &S 68   		     \\
22&05 38 50.78&$-$2 36 26.8&15.06&13.112&0.026&12.445&0.027&12.200&0.025&12.180&0.012&11.800&0.134      &K 9 		     \\
23&05 39 49.45&$-$2 23 45.9&15.15&13.404&0.030&12.758&0.030&12.438&0.030&12.300&0.013&11.847&0.047      &SOri\,J053949.3$-$022346\\ 
24&05 39 20.97&$-$2 30 33.5&15.40&13.286&0.027&12.753&0.027&12.438&0.029&12.359&0.013&11.959&0.043      &SOri\,3     \\ 
25&05 39 05.24&$-$2 33 00.6&14.97&13.394&0.028&12.720&0.024&12.462&0.027&12.500&0.016&12.269&0.034      &K 5  		     \\
26&05 38 23.58&$-$2 20 47.6&15.24&13.412&0.026&12.799&0.029&12.490&0.031&12.452&0.014&11.866&0.077      &K 14  		     \\
27&05 38 44.49&$-$2 40 30.5&14.98&13.365&0.034&12.724&0.033&12.497&0.035&12.418&0.016&12.095&0.042      &K 6  		     \\
28&05 37 52.11&$-$2 56 55.2&15.26&13.395&0.028&12.826&0.024&12.515&0.025&12.477&0.017&12.009&0.059      &K 15  		     \\
29&05 39 39.32&$-$2 32 25.3&15.48&13.435&0.027&12.910&0.023&12.526&0.027&12.497&0.016&12.075&0.030      &SOri\,4\\ 
30&05 38 23.32&$-$2 44 14.2&15.27&13.462&0.027&12.852&0.023&12.562&0.024&12.550&0.017&12.321&0.025      &K 16  		     \\
31&05 38 47.66&$-$2 30 37.4&15.33&13.449&0.030&12.847&0.026&12.585&0.025&12.512&0.015&12.152&0.025      &SOri\,6   \\ 
32&05 39 01.16&$-$2 36 38.9&15.11&13.522&0.027&12.895&0.027&12.605&0.027&12.593&0.014&12.236&0.111      &K 10  		     \\
33&05 38 16.10&$-$2 38 04.9&15.17&13.583&0.027&12.878&0.023&12.612&0.032&12.589&0.017&12.267&0.066      &K 12  		     \\
34&05 36 46.91&$-$2 33 28.3&15.28&13.547&0.024&12.968&0.030&12.660&0.028&12.604&0.015&12.479&0.031      &K 17  		     \\
35&05 37 15.16&$-$2 42 01.6&15.07$^{*}$&13.617&0.029&12.996&0.031&12.776&0.030&12.670&0.015&12.499&0.220&SOri\,J053715.1$-$024202\\
36&05 39 50.57&$-$2 34 13.7&15.39&13.677&0.030&13.002&0.026&12.732&0.027&12.682&0.015&12.308&0.074      &K 20  		     \\
37&05 38 54.92&$-$2 28 58.3&15.44&13.804&0.030&13.201&0.026&12.865&0.030&12.916&0.020&12.634&0.086      &K 22  		     \\
38&05 39 08.22&$-$2 32 28.4&15.71&13.798&0.026&13.254&0.026&12.917&0.029&12.895&0.021&12.590&0.018      &SOri\,7  \\
39&05 38 50.61&$-$2 42 42.9&15.90&13.843&0.030&13.246&0.026&12.963&0.038&12.882&0.021&12.291&0.054      &K 32  		     \\
40&05 39 43.00&$-$2 13 33.3&15.53&13.901&0.027&13.278&0.026&12.990&0.024&13.010&0.041&12.609&0.047      &K 25  		     \\
41&05 37 56.14&$-$2 09 26.7&15.68&13.905&0.027&13.291&0.027&13.035&0.026&12.945&0.017&12.352&0.100      &K 28  		     \\
42&05 37 50.32&$-$2 12 24.7&15.56&13.898&0.024&13.305&0.029&13.044&0.024&13.036&0.019&12.540&0.069      &K 26  		     \\
43&05 39 44.51&$-$2 24 43.2&16.30&14.167&0.029&13.539&0.024&13.150&0.032&13.066&0.021&12.629&0.030      &SOri\,10	     \\ 
44&05 38 49.29&$-$2 23 57.6&16.27&14.362&0.027&13.699&0.026&13.197&0.030&13.194&0.024&12.270&0.090      &SOri\,J053849.2$-$022358\\ 
45&05 37 57.46&$-$2 38 44.4&16.28&14.226&0.030&13.634&0.029&13.285&0.033&13.286&0.027&12.802&0.016      &SOri\,12	     \\ 
46&05 38 16.99&$-$2 14 46.3&15.97&14.237&0.027&13.661&0.032&13.346&0.033&13.274&0.027&12.914&0.051      &K 33  		     \\
47&05 38 10.12&$-$2 54 50.7&16.47&14.391&0.027&13.757&0.032&13.425&0.045&13.405&0.031&12.773&0.023      &K 43  		     \\
48&05 39 37.60&$-$2 44 30.5&16.75&14.380&0.031&13.819&0.027&13.384&0.034&13.350&0.029&13.450&0.057      &SOri\,14		     \\
49&05 38 48.10&$-$2 28 53.6&16.39&14.470&0.033&13.840&0.026&13.435&0.036&13.453&0.030&12.915&0.060      &SOri\,15	     \\   
50&05 38 47.15&$-$2 57 55.7&17.05&14.515&0.032&13.935&0.036&13.461&0.044&13.532&0.027&12.812&0.098      &K 56  		     \\
51&05 39 11.40&$-$2 33 32.8&16.42&14.452&0.034&13.929&0.029&13.571&0.043&13.560&0.026&13.334&0.115      &SOri\,J053911.4$-$023333\\
52&05 38 13.31&$-$2 51 33.0&16.62&14.570&0.034&13.996&0.033&13.636&0.045&13.627&0.029&13.022&0.042      &K 48  		     \\
53&05 38 38.59&$-$2 41 55.9&16.47&14.562&0.031&13.972&0.030&13.665&0.040&13.647&0.029&13.256&0.018      &K 44  		     \\
54&05 40 32.49&$-$2 40 59.8&16.97&14.703&0.038&14.065&0.032&13.709&0.049&13.661&0.031&13.271&0.079      &K 54  		     \\
55&05 38 35.36&$-$2 25 22.2&16.88&14.652&0.033&14.056&0.036&13.764&0.041&13.725&0.035&13.264&0.039      &SOri\,22     \\
56&05 39 34.33&$-$2 38 46.9&17.25&14.763&0.032&14.188&0.037&13.787&0.050&13.755&0.034&13.053&0.031      &SOri\,21	     \\ 
57&05 38 17.42&$-$2 40 24.3&17.20&14.833&0.031&14.314&0.037&14.093&0.054&13.926&0.043&13.505&0.160      &SOri\,27	     \\   
58&05 38 51.00&$-$2 49 14.0&16.79&15.040&0.039&14.421&0.033&14.159&0.070&14.140&0.048&13.752&0.091      &K 50  		     \\
59&05 37 27.62&$-$2 57 10.0&17.08&15.122&0.052&14.480&0.065&14.310&0.088&14.224&0.050&13.662&0.240      &K 57  		     \\
\hline
\end{tabular}
\end{minipage}
\end{table*}

\begin{figure*}[h]
\includegraphics[height=16cm]{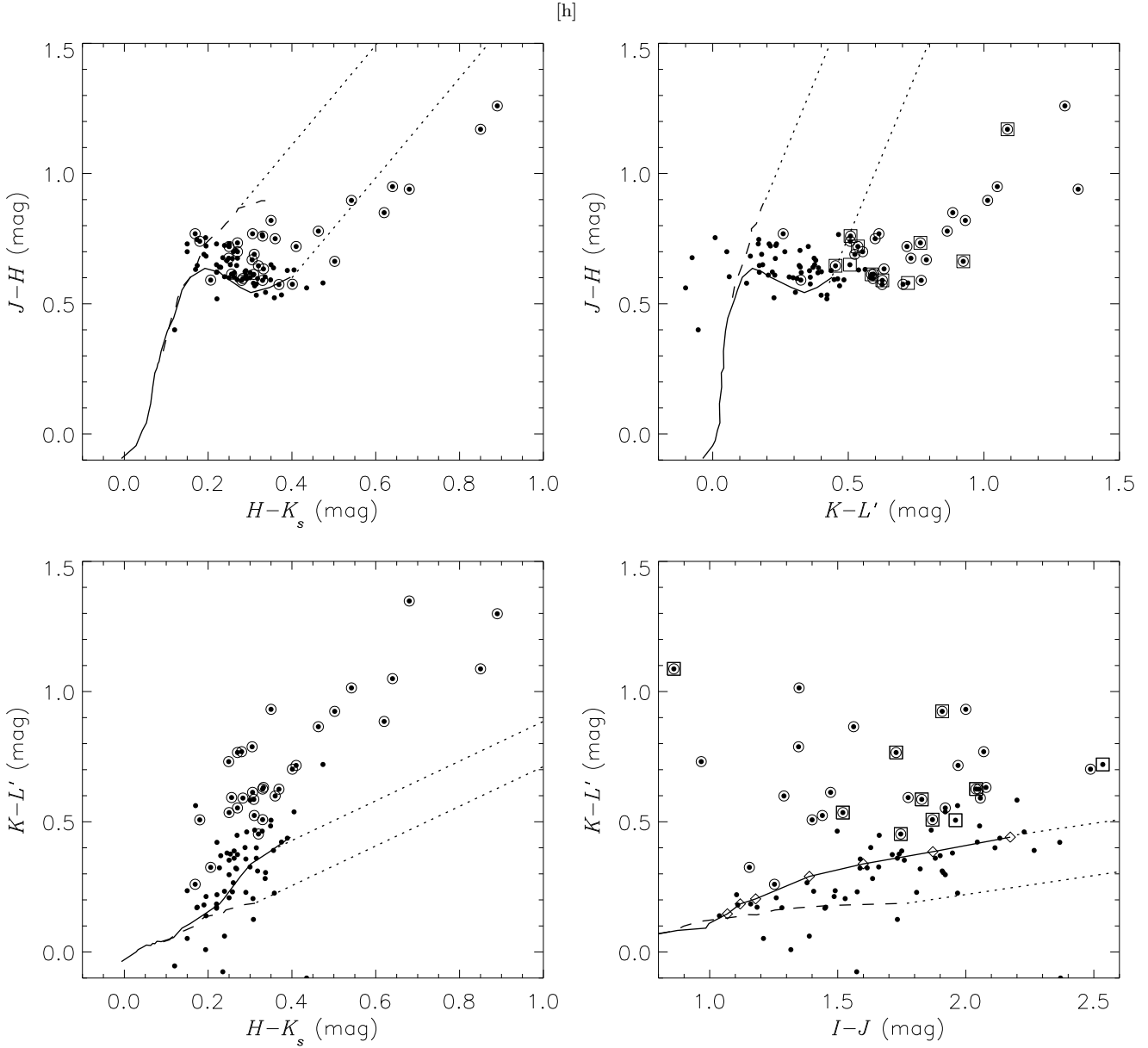}
\caption{Colour-colour diagrams for the 87 $\sigma$\,Ori cluster members; from 
left to right and top to bottom: $JHK_{\rm s}$ from 2MASS, $JHKL'$, 
$HK_{\rm s}KL'$ and $IJKL'$ from 2MASS and our UKIRT observations. The solid and
dashed lines are respectively the empirical loci for main-sequence (to spectral
type M6) and giant stars and the dotted lines are reddening bands (see text). 
Double-circle objects have 2-$\sigma$ $(K-L')$ excess detections. In the $JHKL'$
and $IJKL'$ diagrams we indicate (open squares) objects with H$\alpha$ 
equivalent width indicative of stellar accretion (Sect.\,\ref{acc}). Open 
diamonds in the $IJKL'$ diagram show the colours of main-sequence stars with
spectral types M0$-$M6.}
\label{ccd}
\end{figure*} 

The targets' positions, $I_{\rm c}$ magnitudes, 2MASS  $J, H$ and $K_{\rm s}$ 
photometry, the new K- and L$'$-band measurements and identifications are listed
in Table\,\ref{obs_table}. $I_{\rm c}$ magnitudes are mostly from optical 
observations described in \citet{kenyon05}. The 6 objects indicated by $^{*}$ in
Table\,\ref{obs_table} were either saturated or out of their survey area; in 
these cases we used photometry from \citet{sherry04} and \citet{osorio02}. In
the analysis described here, we merge the present sample (59 objects) with the 
sample from \citet[][  28 objects]{oliveira04b} as the method for object 
selection is identical. The total sample thus comprises 87 objects although the 
4 known IRAS sources from \citet{oliveira04b} are excluded from the discussion 
of the IR excesses and disc frequency, as their selection criteria were 
different from the remainder of the sample.

\section{Analysis of the $(K-L')$ colours of the cluster members}

A $(K-L)$ excess in a young star has been proven to be a reliable disc 
diagnostic \citep{haisch01,wood02}. Determining whether an object has such an
excess can be done in two ways: by using colour-colour diagrams, or, if the 
object's spectral type is known, by directly computing the excess above the 
photospheric emission.

\subsection{Colour-colour diagrams}
\label{excess}

Fig.\,\ref{ccd} shows several colour-colour diagrams for all 87 cluster members.
Each diagram shows the object's colours with respect to the loci of 
main-sequence and giant stars  --- loci from \citet{bessell88} are converted to
the appropriate photometric systems using transformations from 
\citet{hawarden01} and \citet{carpenter01}. Objects have an excess in 
$(H-K_{\rm s})$ or $(K-L')$ colour when they appear to the right of the 
reddening band ($JHK_{\rm s}$ and $JHKL'$ diagrams, top of Fig.\,\ref{ccd}). Few
objects show an excess in $(H-K_{\rm s})$, as expected since only accreting 
discs with high-mass accretion rates have dust warm enough to produce an excess
in the K-band (Sect.\,\ref{acc}). In the $JHKL'$diagram, 33 objects out of the 
83 objects appear to the right of the reddening band, indicating a disc 
frequency of $\sim$\,(40\,$\pm$\,7)\%. Early type objects with small excesses 
may remain undetected. The computed disc frequency also depends on the adopted 
spectral type boundary, as noted by \citet{lada04}. In the next sections we 
compute $(K-L')$ excesses and identify objects with circumstellar discs, taking
into account the corresponding uncertainties.

Fig.\,\ref{ccd} (bottom right) also shows the $IJKL'$ colour-colour diagram 
for the target sample. In this diagram the giant locus (dashed line) sits below 
the main-sequence locus (solid line). This diagram offers a very clean method of
identifying objects with circumstellar discs (i.e. objects that appear above the
main-sequence locus) and it is free of some of the problems that affect the 
usage of other colour-colour diagrams. In particular for late spectral types, 
the main-sequence locus is for all effects horizontal, meaning that 
uncertainties in a star's $(I-J)$ colour (e.g., due to accretion variability or
stellar activity) do not significantly affect the identification of an $(K-L')$
excess. Furthermore, the effect of reddening is to move a star's position 
approximately along the same locus, again having a small impact in the disc 
analysis. This diagram is also more sensitive to early type objects with small 
excesses and does not depend on any ad hoc spectral-type boundary. Using this 
diagram, the $(K-L')$ disc frequency for $\sigma$\,Ori cluster members is 
$\sim$\,(50\,$\pm$\,7)\%, higher than determinations using the $JHKL'$ 
colour-colour diagram and measured $(K-L')$ excesses (see next section). It is 
clear that the derived disc frequency is somewhat dependent on the method used.

\subsection{$(K-L')$ excesses}

\begin{table*}
\centering
\begin{minipage}{180mm}
\caption{IR excesses for the target sample. Column 2 gives the spectral type; 
{$\dag$} indicates objects with spectral types estimated from their colours. 
Columns 3, 4 and 5 and 6, 7 and 8 are the measured colours, excesses and 
uncertainties, respectively for $(H-K_{\rm s})$ and $(K-L')$. Column 9 is the 
H$\alpha$ equivalent width, EW[H$\alpha$]. Column 10 gives references for the 
spectral type classification and/or EW[H$\alpha$]: \citet[][ ZO02]{osorio02}, 
\citet[][ B03]{barrado03}, \citet[][ SE04]{scholz04} 
and \citet[][ M03]{muzerolle03}. ``a'' in column 9 indicates accreting objects, 
based on H$\alpha$ emission line width \citep[][ K05]{kenyon05}. Column 11 is 
the objects' identifications.}
\label{excess_table}
\begin{center}
\begin{tabular}{llllllllccl}
\hline
  & SpT  & \multicolumn{3}{c}{$(H-K_{\rm s})$}   & \multicolumn{3}{c}{$(K-L')$} & \multicolumn{1}{c}{EW[H$\alpha$]} &\multicolumn{1}{l}{References}&\multicolumn{1}{l}{Identification}\\
  &      &   \llap{o}bserved     & excess    & \llap{e}rror            & observed & excess     &\llap{e}rror              & (\AA)          &\\
\hline
 1& M3$\dag$   &0.542& 0.265&0.052& 1.014& 0.724&0.072&                            &       &S 33	                \\
 2& K6         &0.174& 0.026&0.038& 0.172& 0.071&0.042&2\rlap{.2}                  &ZO02   &4771$-$1097                 \\
 3& K7.5       &0.249& 0.074&0.039& 0.731& 0.603&0.041&6\rlap{.4}                  &ZO02   &4771$-$1051                 \\
 4& M3$\dag$   &0.305& 0.028&0.050& 0.788& 0.497&0.066&                            &       &S 145	                \\
 5& M4$\dag$   &0.463& 0.161&0.071& 0.865& 0.527&0.066&                            &       &S 102		        \\
 6& M1$\dag$   &0.206&\llap{$-$}0.021&0.046& 0.345& 0.140&0.046&                   &       &S 166		        \\
 7& M0$\dag$   &0.195& 0.002&0.044& 0.139&\llap{$-$}0.007&0.081&                   &       &S 99		        \\
 8& M1.5$\dag$ &0.195&\llap{$-$}0.037&0.043& 0.213& 0.018&0.067&                   &       &S 125		        \\
 9& M0.5$\dag$ &0.173&\llap{$-$}0.036&0.044& 0.170& 0.004&0.045&                   &       &S 61		        \\
10& M4$\dag$   &0.329& 0.027&0.049& 0.464& 0.125&0.065&                            &       &S 41		        \\
11& M4         &0.253&\llap{$-$}0.048&0.041& 0.376& 0.037&0.039&\llap{1}4          &ZO02   &r053833$-$0236              \\
12& M0.5$\dag$ &0.169&\llap{$-$}0.040&0.044& 0.260& 0.094&0.043&                   &       &S 138		        \\
13& M1$\dag$   &0.194&\llap{$-$}0.033&0.048& 0.009&\llap{$-$}0.175&0.076&          &       &S 143 	                \\
14& M1$\dag$   &0.221&\llap{$-$}0.006&0.044& 0.220& 0.035&0.059&                   &       &S 168		        \\
15& M2.5$\dag$ &0.239&\llap{$-$}0.018&0.051& 0.233&\llap{$-$}0.014&0.113&          &       &S 46		        \\
16& M4$\dag$   &0.268&\llap{$-$}0.033&0.055& 0.319&\llap{$-$}0.019&0.081&          &       &S 16		        \\
17& M4         &0.270&\llap{$-$}0.031&0.035& 0.766& 0.427&0.038&\llap{2}8          &ZO02   &r053820$-$0234              \\
18& M4$\dag$   &0.307& 0.005&0.052& 0.205&\llap{$-$}0.133&0.107&                   &       &S 28		        \\
19& M3.5$\dag$ &0.306& 0.016&0.050& 0.613& 0.298&0.127&                            &       &S 39		        \\
20& M2.5$\dag$   &0.239& \llap{$-$}0.018&0.053& 0.061&\llap{$-$}0.186&0.078&       &       &S 38		        \\
21& M3$\dag$ &0.235&\llap{$-$}0.042&0.052&\llap{$-$}0.076&\llap{$-$}0.366&0.209&   &       &S 68		        \\
22& M4$\dag$   &0.245&\llap{$-$}0.056&0.054& 0.380& 0.041&0.147&                   &       &K 9 		        \\
23& M4         &0.320& 0.018&0.046& 0.453& 0.114&0.057&\llap{4}2                   &ZO02   &SOri\,J053949.3$-$022346    \\
24& M5$\dag$   &0.315&\llap{$-$}0.031&0.056& 0.400& 0.014&0.074&                   &       &SOri\,3	   	        \\
25& M3$\dag$   &0.258&\llap{$-$}0.018&0.053& 0.231&\llap{$-$}0.059&0.070&          &       &K 5 		        \\
26& M4.5$\dag$ &0.309&\llap{$-$}0.015&0.058& 0.586& 0.224&0.098&\rlap{a}           &K05    &K 14		        \\
27& M2.5$\dag$ &0.227&\llap{$-$}0.030&0.062& 0.323& 0.075&0.074&                   &       &K 6 		        \\
28& M4.5$\dag$ &0.311&\llap{$-$}0.013&0.052& 0.468& 0.106&0.085&                   &       &K 15		        \\
29& M5.5$\dag$   &0.375& 0.003&0.053& 0.422& 0.008&0.068&                          &       &SOri\,4   		        \\
30& M4$\dag$   &0.290&\llap{$-$}0.011&0.052& 0.229&\llap{$-$}0.109&0.067&          &       &K 16		        \\
31& M4$\dag$   &0.262&\llap{$-$}0.039&0.053& 0.360& 0.021&0.066&                   &       &SOri\,6      	        \\
32& M4$\dag$   &0.290&\llap{$-$}0.011&0.055& 0.357& 0.018&0.126&                   &       &K 10		        \\
33& M3$\dag$   &0.266&\llap{$-$}0.010&0.056& 0.322& 0.031&0.090&                   &       &K 12		        \\
34& M4.5$\dag$ &0.308&\llap{$-$}0.016&0.057& 0.125&\llap{$-$}0.236&0.072&          &       &K 17		        \\
35& M4         &0.220&\llap{$-$}0.081&0.047& 0.172&\llap{$-$}0.166&0.223&4\rlap{.9}&ZO02   &SOri\,J053715.1$-$024202    \\
36& M3.5$\dag$ &0.270&\llap{$-$}0.019&0.054& 0.374& 0.053&0.099&                   &       &K 20		        \\
37& M4$\dag$   &0.336& 0.034&0.056& 0.282&\llap{$-$}0.056&0.106&                   &       &K 22		        \\
38& M5$\dag$   &0.337&\llap{$-$}0.009&0.055& 0.305&\llap{$-$}0.080&0.066&          &       &SOri\,7     	        \\
39& M4$\dag$   &0.283&\llap{$-$}0.018&0.060& 0.591& 0.252&0.083&                   &       &K 32		        \\
40& M4$\dag$   &0.288&\llap{$-$}0.013&0.053& 0.401& 0.062&0.086&                   &       &K 25		        \\
41& M3.5$\dag$ &0.256&\llap{$-$}0.033&0.054& 0.593& 0.278&0.117&                   &       &K 28		        \\
42& M3.5$\dag$ &0.269&\llap{$-$}0.020&0.054& 0.448& 0.133&0.093&                   &       &K 26		        \\
43& M6$\dag$   &0.389&\llap{$-$}0.007&0.056& 0.437&\llap{$-$}0.003&0.070&          &       &SOri\,10		        \\
44& M5$\dag$   &0.502& 0.155&0.056& 0.924& 0.538&0.110&\llap{2}1\rlap{.3a}         &SE04, K05&SOri\,J053849.2$-$022358   \\
45& M6         &0.349&\llap{$-$}0.047&0.048& 0.484& 0.043&0.043&9                  &M03    &SOri\,12		        \\
46& M4.5$\dag$ &0.315&\llap{$-$}0.009&0.060& 0.360&\llap{$-$}0.001&0.083&          &       &K 33		        \\
47& M5$\dag$   &0.332&\llap{$-$}0.014&0.068& 0.632& 0.246&0.071&                   &       &K 43		        \\
48& M6.5$\dag$ &0.435& 0.013&0.059&\llap{$-$}0.100&\llap{$-$}0.568&0.087&          &       &SOri\,14		        \\
49& M5.5       &0.405& 0.033&0.048& 0.538& 0.124&0.073&\llap{1}5\rlap{.7}          &B03    &SOri\,15		        \\
50& M7$\dag$   &0.474& 0.027&0.069& 0.720& 0.223&0.118&\rlap{a}                    &K05    &K 56		        \\
51& M5         &0.358& 0.011&0.055& 0.226&\llap{$-$}0.159&0.121&4\rlap{.7}         &B03    &SOri\,J053911.4$-$023333    \\
52& M5.5$\dag$ &0.369&\llap{$-$}0.002&0.068& 0.625& 0.211&0.078&                   &       &K 48		        \\
53& M5$\dag$   &0.318&\llap{$-$}0.028&0.064& 0.311&\llap{$-$}0.074&0.069&          &       &K 44		        \\
54& M5.5$\dag$ &0.356&\llap{$-$}0.015&0.070& 0.390&\llap{$-$}0.023&0.103&          &       &K 54		        \\
55& M6         &0.292&\llap{$-$}0.104&0.058& 0.461& 0.020&0.060&6\rlap{.8}         &B03    &SOri\,22	   	        \\
56& M6.5$\dag$ &0.401&\llap{$-$}0.020&0.073& 0.702& 0.233&0.075&                   &       &SOri\,21		        \\
57& M7         &0.221&\llap{$-$}0.225&0.068& 0.421&\llap{$-$}0.075&0.168&5\rlap{.1}&B03    &SOri\,27		        \\
58& M4.5$\dag$ &0.262&\llap{$-$}0.062&0.087& 0.388& 0.026&0.119&                   &       &K 50		        \\
59& M5$\dag$   &0.170&\llap{$-$}0.176&0.116& 0.562& 0.176&0.252&                   &       &K 57		        \\
\hline										      
\end{tabular}
\end{center}
\end{minipage}
\end{table*}

\begin{figure}
\includegraphics[height=9.5cm]{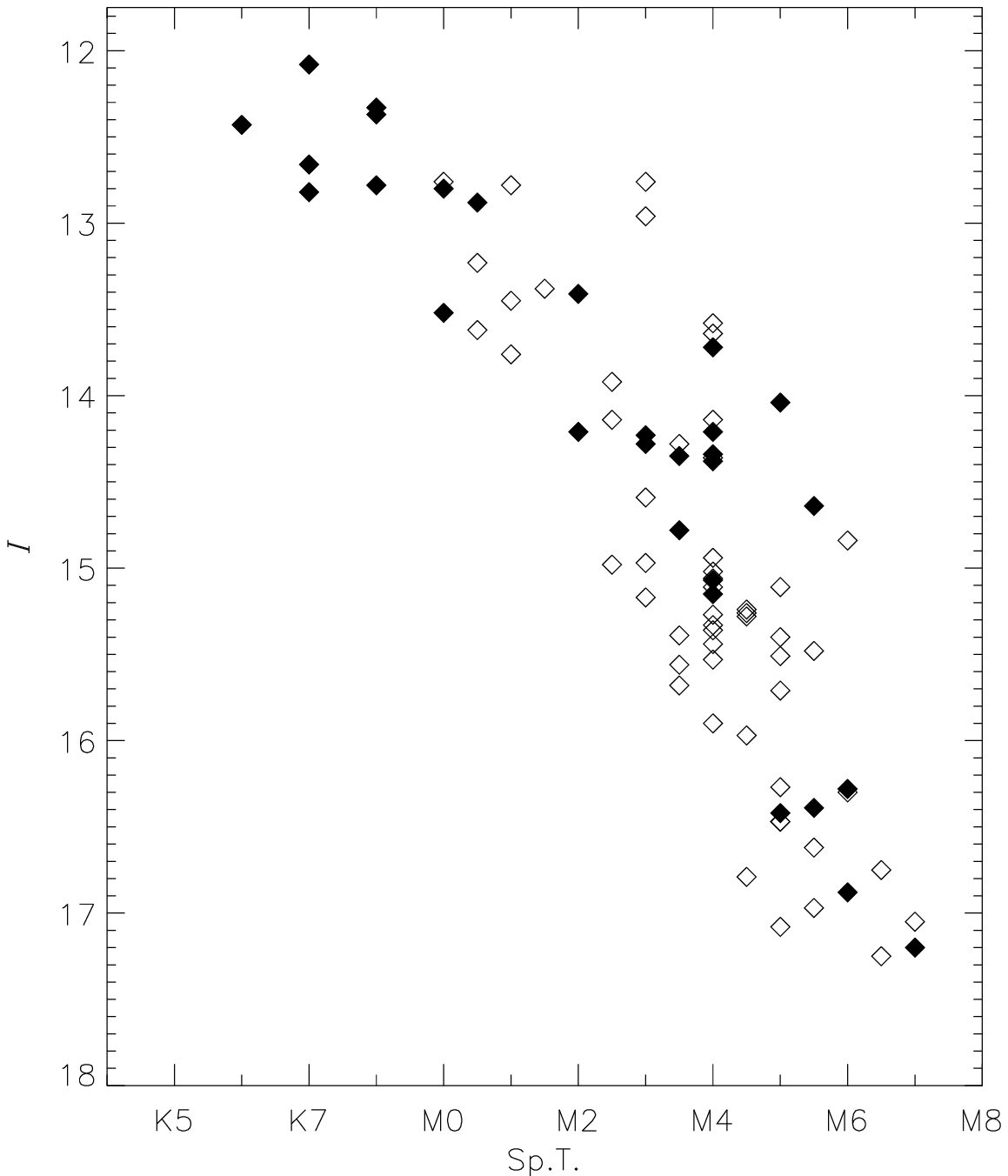}
\caption{Spectral types versus I-band magnitude for observed cluster members.
Filled symbols are objects which have published spectral types, determined 
from optical spectra, while open symbols represent objects with spectral types 
determined from observed colours. The scatter in computed spectral types 
is not significantly larger than the scatter from ``classical'' spectral 
typing.}
\label{spt}
\end{figure}

We can compute $(K-L')$ excesses individually for each young star if spectral 
types are known or can estimated. As it can be seen from 
Table\,\ref{excess_table}, spectral types are only known for a small fraction of
the observed sample. Following \citet{oliveira04b}, for objects without 
published spectral types, we estimate these from the observed colours. We use
tabulated relations between colours and spectral type from \citet{bessell88}, 
converted to the appropriate photometric system. We estimate the spectral type
that corresponds respectively to the observed $(I-J)$ and $(H-K_{\rm s})$ 
colours --- for objects with no obvious $(H-K_{\rm s})$ excess, 
Sect.\,\ref{acc}. These two quantities are then averaged to compute the values 
listed in Table\,\ref{excess_table} (indicated by $\dag$). These spectral type 
determinations are uncertain by about a subclass \citep[see][ for a full discussion]{oliveira04b}; 
spectral types measured from optical spectroscopy typically have quoted 
uncertainties of half a subclass. Accordingly, the excess error listed in the 
table combines in quadrature the photometric errors with this spectral type 
error. Fig.\,\ref{spt} shows the spectral types (both estimated and known from 
the literature) for the sample, plotted against I-band magnitude. A similar 
diagram (effectively a colour-magnitude diagram) can be found in 
\citet{barrado03}. For each spectral type, the ``scatter'' in magnitude is 
similar between objects with published and estimated spectral types, suggesting
reliable spectral type determinations. Both $(I-J)$ and $(H-K_{\rm s})$ 
colours can be affected by the disc itself and interstellar reddening. However,
we find good agreement between spectral types computed by this colour method and 
derived directly from spectroscopy. Furthermore, a mere glance at the $IJKL'$ 
diagram in Fig.\,\ref{ccd} shows that the disc frequency determined in this way 
is not very sensitive to spectral type (e.g., $(I-J)$) uncertainties (see next 
paragraph). Therefore we are confident these spectral type determinations are 
appropriate, {\em as an aide for the disc frequency determination and as long as
spectral type uncertainties are taken into account}. 

\begin{figure*}
\includegraphics[height=8.5cm]{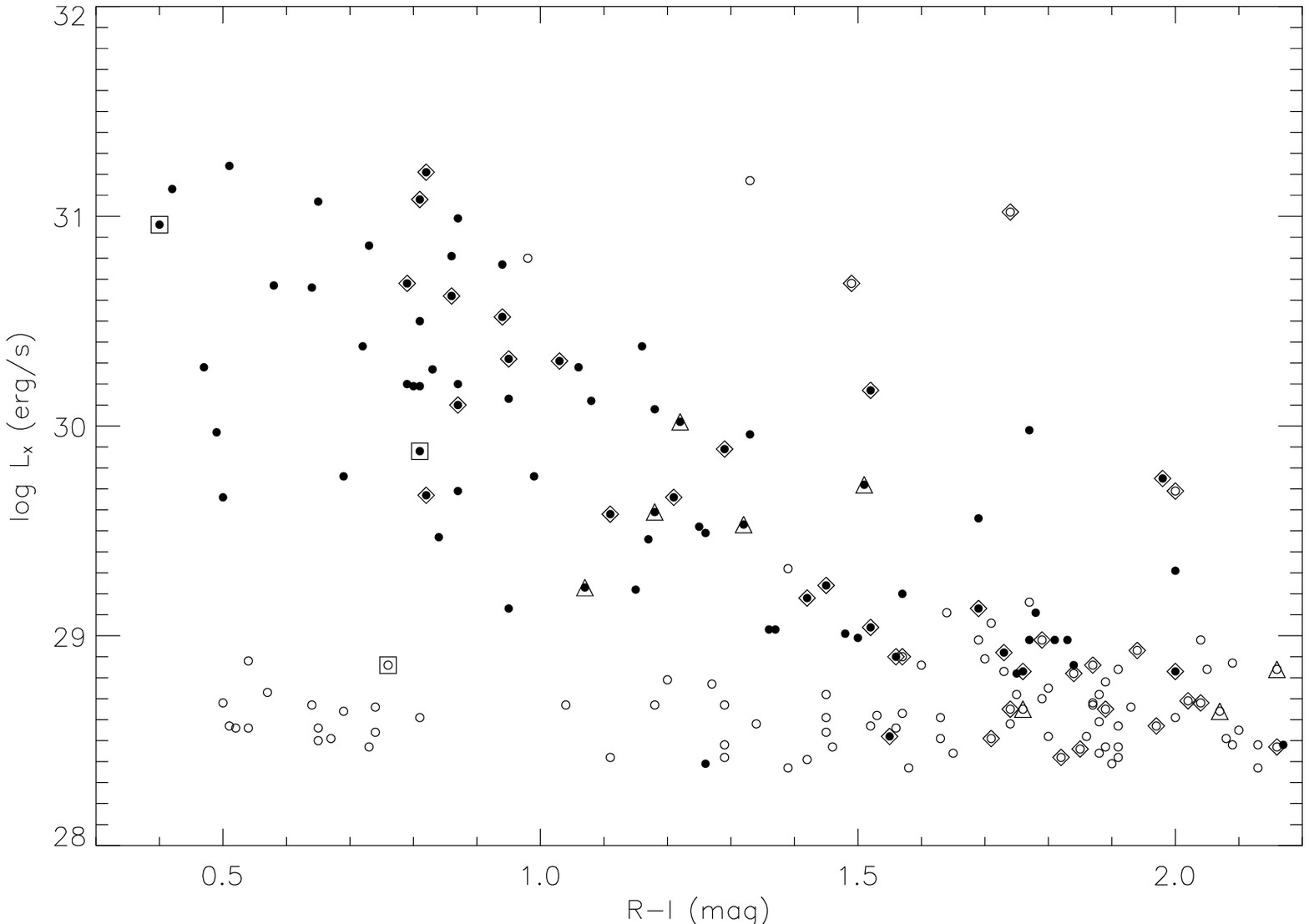}
\caption{X-ray luminosities for $\sigma$\,Ori cluster members and candidates
from \citet{franciosini05}. Filled circles are XMM-{\em Newton} detections and 
open circles are 3-$\sigma$ upper limits (see their Tables A1 and A2). Also 
plotted are the objects in our sample that fall onto the EPIC field: diamonds 
are spectroscopic cluster members, triangles are photometric candidates and 
squares are IRAS sources (see text).}
\label{lx}
\end{figure*} 

Of the 83 cluster members in the sample (IRAS sources excluded), 27 objects have
2-$\sigma$ significant $(K-L')$ excesses, indicating a disc frequency of 
(33\,$\pm$\,6)\%. This value is consistent, at the 1-$\sigma$ level, with 
(46\,$\pm$\,14)\% computed by \citet{oliveira04b} for a subset of 24 objects
(but see Sect.\,\ref{mass}). These figures are conservative, as we demand a 
2-$\sigma$ significant detection as a disc indicator. Furthermore, for fainter 
targets increasing L$'$-band uncertainties mean that we might be unable to 
detect a small excess. The analysis of the $IJKL'$ diagram in Fig.\,\ref{ccd} 
confirms that all 2-$\sigma$ detections are reliable and it also suggests that 
more objects might have a $(K-L')$ excess, just not statistically significant. 
Thus this disc frequency should be taken as a lower limit. 

\section{Discussion}

\subsection{Sample contamination and X-ray luminosities}

Most of our sample has supporting spectroscopic evidence of cluster membership 
(i.e. their spectra show spectroscopic indicators of youth and their radial 
velocities are consistent with cluster membership, see 
Sect.\,\ref{target_sample}) but 19 objects are selected only on the basis of 
their photometry. Because contaminating field stars would be unlikely to have 
circumstellar emission it is important to assess the level of this contamination
and determine its effect on our derived disc frequencies. As young stars are (on
average) much more X-ray luminous than any older contaminating field objects 
\citep[e.g.,][]{feigelson03}, we can use X-ray luminosities as another 
membership indicator.

\citet{franciosini05} have recently analysed XMM-{\em Newton} observations of 
an EPIC field centered on the O-star $\sigma$\,Ori (area of diameter 
$\sim$\,30\,arcmin). From our sample, 51 targets lie within the EPIC field of 
view. Cross-correlating these targets with the X-ray data we find 30 objects 
with X-ray detections and 3-$\sigma$ upper limits for the rest. 

The X-ray luminosities are presented as a function of $(R-I)$ colour in
Fig.\,\ref{lx}. Of the 19 photometric candidates, 8 are in the EPIC field and 5
are detected with X-ray luminosities greater than $10^{29}$\,erg\,s$^{-1}$. It 
is highly unlikely that contaminating field M-dwarfs would have X-ray 
luminosities as high as this \citep[see][ and references therein for an analysis of X-ray luminosity evolution with time]{jeffries05}, 
and so these 5 young M stars are highly probable cluster members. The 3 
non-detections are at colours of $(R-I)>1.7$ where the lack of an X-ray 
detection does not adequately discriminate between very young stars and field
M-dwarfs because the X-ray survey is not sensitive enough. Indeed, many 
spectroscopically confirmed members are not detected in X-rays at these colours
(Fig.\,\ref{lx}). According to \citet{kenyon05}, almost all objects in 
photometrically selected cluster samples with $(R-I)\,>\,1.7$ are revealed as
genuine cluster members when spectroscopic indicators are used, so we have 
little doubt that the contamination among these cool photometrically selected 
candidates is very small.

Even if we were reasonably conservative and use the published photometric 
membership probabilities of \citet{sherry04} --- i.e. ignoring the X-ray 
evidence --- we estimate that 17\% of the 19 photometric candidates might be 
contaminants. Even if these $\sim 3$ objects are assumed not to have discs and 
are removed from the sample it cannot raise the deduced disc frequency 
significantly. Alternatively, we can simply ignore the 11 photometric cluster 
candidates that lie outside the EPIC field of view and therefore have no 
supporting X-ray evidence for their membership. The resulting disc frequency is
still 32\%, consistent with our estimates in the previous section. We have also
confirmed that such a procedure would not affect our conclusions regarding any 
mass dependence of the disc frequency (see next section).

The disc frequency for objects in the EPIC field is (33\,$\pm$\,8)\% while the
disc frequency for objects outside this central area is (31\,$\pm$\,9)\%. These
estimates are indistinguishable from the total disc frequency of 
(33\,$\pm$\,6)\%, thus we find no evidence for any spatial segregation of 
objects with and without circumstellar discs.

\subsection{Mass dependence of disc frequency}
\label{mass}

\begin{figure*}
\includegraphics[height=8.cm]{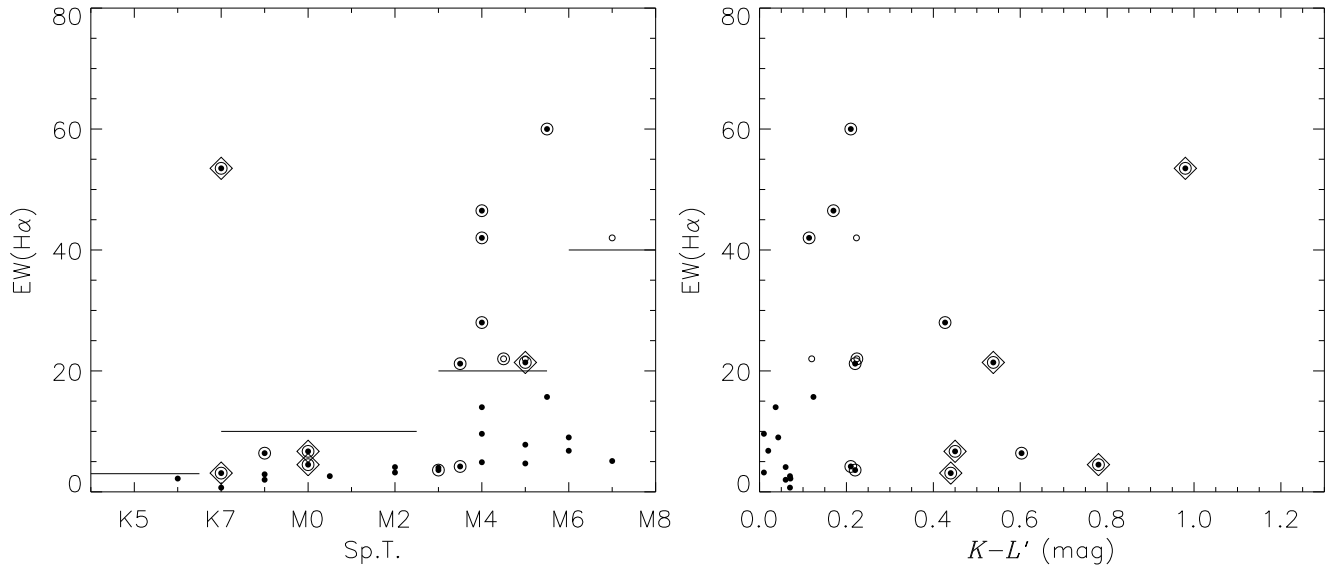}
\caption{H$\alpha$ measurements as a function of spectral type and $(K-L')$ 
colour. Objects either have EW[H$\alpha$] measurements in 
Table\,\ref{excess_table} (filled circles) or have been identified as accretors
based on the line width \citep[][ open circles]{kenyon05}. Double-circle 
objects have 2-$\sigma$ significant $(K-L')$ excesses while diamonds are 
objects with 2-$\sigma$ significant $(H-K_{\rm s})$ excesses (see text). The 
full lines represent the \citet{white03} criterion for the identification of 
accreting cluster members.}
\label{halpha}
\end{figure*} 

One of the main goals behind the $\sigma$\,Ori cluster survey was to 
investigate the possibility of a mass dependence in the disc frequency. Using 
the colour-magnitude diagram and isochrones from Fig.\,\ref{cmd}, we have 
determined the masses and ages of the cluster members. We have divided the 
sample in 3 mass bins and determined the disc frequency in each bin. The mass 
bins are 
M\,$\geq$\,0.5\,M$_{\odot}$, 0.5\,M$_{\odot}$\,$>$\,M\,$\geq$\,0.1\,M$_{\odot}$ 
and 0.1\,M$_{\odot}$\,$>$\,M\,$\geq$\,0.04\,M$_{\odot}$; the respective disc
frequencies are 11/26 or (42\,$\pm$\,13)\%, 13/45 or (28\,$\pm$\,8)\% and
3/12 or (25\,$\pm$\,14)\%. Thus, there is a hint of a decrease in disc frequency
towards lower masses that could also explain the slightly higher disc frequency 
determined by \citet{oliveira04b}. However this result is not conclusive and 
these disc frequencies are entirely consistent with no mass dependence. In 
particular in the lower mass bin, objects have redder intrinsic colours and 
with increasing uncertainties in the L$'$-band photometry, the measurements are
less capable of identifying excesses.

It is still not clear whether disc frequencies or disc dissipation time scales 
are mass dependent. \citet{lada00} find high disc frequencies from F to late M 
spectral types for the 1\,Myr-old Trapezium cluster, suggesting that the 
probability of disc formation around a star is both high and essentially mass 
independent. However, in their more recent work, \citet{lada04} find a hint that
the disc frequency around brown dwarfs in the Trapezium cluster might be lower 
than around stellar objects, but completeness issues make this result 
inconclusive. For IC\,348 (2$-$3\,Myr) \citet{haisch01} suggest a disc frequency
decrease from spectral types G down to M. Recently, \citet{lada05} analysed 
Spitzer near- and mid-IR photometry of IC\,348 and they find that the disc 
frequency peaks for late-K to early-M spectral types, suggesting that disc 
dissipation is mass dependent. The disc frequencies we have determined for the
different mass bins are consistent with the IC\,348 result but are also 
statistically consistent with a single disc frequency for the $\sigma$\,Orionis 
cluster across the mass range. To settle this question in a conclusive way, it 
would be necessary to gather a larger sample of very low-mass cluster members 
with better signal-to-noise ratio measurements in the L-band.

\subsection{Accretors in the $\sigma$\,Ori cluster}
\label{acc}

Strong H$\alpha$ emission and broad line profiles were the classical indicator
of young stars with discs, but we now know that they only identify actively
accreting discs \citep[e.g.,][]{white03,muzerolle03,barrado03}. In our sample, 
33 objects have H$\alpha$ equivalent width measurements (EW[H$\alpha$]) or have
been identified as accretors based on the width of the H$\alpha$ profile 
\citep{kenyon05} (Table\,\ref{excess_table} and Fig.\,\ref{halpha}). Using the 
\citet{white03} criterion of EW[H$\alpha$] as a function of spectral type, we 
can identify stars with a H$\alpha$ ``excess emission'', i.e.\, objects that 
exhibit emission strong enough to indicate active accretion onto the star --- 
the \citet{barrado03b} criterion could also be used, yielding the same results.
Out of these 33 cluster members, 10 stars have accretion discs; a fraction of 
$(30 \pm 10)$\% consistent with previous results (30$-$40\%, \citealt{osorio02} 
and $(27 \pm 7)$\%, \citealt{barrado03}). Of the 10 objects with an accretion 
disc, 8 objects have 2-$\sigma$ significant $(K-L')$ disc detections, while 2 
fainter objects have $(K-L')$ excesses at a significance level $>$\,1.5-$\sigma$
(marginal disc detections). Thus, probably all H$\alpha$-identified accretors 
have circumstellar discs indicated by a $(K-L')$ excess. However, 6 objects show
a $(K-L')$ excess but no H$\alpha$ excess emission, i.e.\ as expected not all 
objects with discs are actively accreting.

Of the total sample of 87 objects, 11 stars or about $(13 \pm 4)$\,\% show a 
2-$\sigma$  $(H-K_{\rm s})$ excess, also considered an accretion indicator 
\citep{hillenbrand98} --- this is consistent with previous determinations, 
$(6 \pm 4)$\,\% from \citet{oliveira02} and 5$-$12\% from \citet{barrado03}. Of 
the 33 objects with H$\alpha$ measurements, 5 objects show an excess in the 
K$_{\rm s}$-band. We note that not all objects with $(H-K_{\rm s})$ excess show 
H$\alpha$ excess emission and vice-versa (Fig.\,\ref{halpha}), a fact that can
be understood in terms of variability, either in the H$\alpha$\ profile 
\citep{guenther97} or in the K$_{\rm s}$-band magnitudes 
\citep{carpenteretal01}.

The comparison of these indicators reinforces the ``hierarchy'' of disc 
identifiers. In a sample of young objects with measurements of all three disc
indicators: a $(K-L')$ excess identifies most (if not all) circumstellar discs, 
H$\alpha$ excess emission identifies those discs actively accreting, while 
$(H-K)$ excesses identify discs with higher accretion rates, with factors like 
system geometry and stellar parameters also playing a role 
\citep{hillenbrand98}.
 
\section{Summary}

We present new K and L$'$ observations of PMS stars in the $\sigma$\,Orionis 
cluster, obtained at UKIRT. We have computed the disc frequency of the cluster 
using a 2-$\sigma$  $(K-L')$ excess as the disc indicator. Of the 83 cluster 
members in the sample, 27 objects have a circumstellar disc, indicating a disc 
frequency of (33\,$\pm$\,6)\%, consistent with previous determinations 
\citep{oliveira04b}. If instead the $JHKL'$ colour-colour diagram is used to 
identify discs, the disc frequency would be (40\,$\pm$\,7)\%. With an age of  
3$-$5\,Myr, this disc frequency is consistent with the 3\,Myr disc half-life as
determined by \citet{haisch01b}. As the age of the cluster is uncertain (values 
between 2.5\,Myr and 8\,Myr can be found in the literature) we are unable to 
derive firmer conclusions on the disc dissipation timescale, but an older age 
for the $\sigma$\,Ori cluster would imply a slower rate of disc dispersal. We 
find no evidence of spatial segregation of objects with or without circumstellar
discs.

We have investigated a possible mass dependence of the disc frequency. We find
that for stars more massive than 0.5\,M$_{\odot}$ the disc frequency is 
(42\,$\pm$\,13)\%, decreasing to (28\,$\pm$\,8)\% and (25\,$\pm$\,14)\%
respectively for stars with masses in the range
0.5\,M$_{\odot}$\,$>$\,M\,$\geq$\,0.1\,M$_{\odot}$ and 
0.1\,M$_{\odot}$\,$>$\,M\,$\geq$\,0.04\,M$_{\odot}$. These different disc
frequencies are statistically consistent with a single disc frequency across
the mass range. On the other hand, it could also hint at a mass dependence
of the disc frequency in the same sense as that found by \citet{lada05} in 
IC\,348.

For some cluster members in our sample, H$\alpha$\ equivalent width measurements
are also available. We investigate the hierarchy of disc indicators and confirm
that a $(K-L')$ excess identifies the majority of circumstellar discs, H$\alpha$
excess emission identifies discs actively accreting, while a $(H-K)$ excess 
identifies discs with higher accretion rates primarily around the more massive 
young stars. 

\section*{Acknowledgements}
We thank the staff of United Kingdom Infra-Red Telescope (UKIRT) for their 
support. UKIRT is operated by the Joint Astronomy Centre on behalf of the UK 
Particle Physics and Astronomy Research Council (PPARC). This publication makes
use of data products from the Two Micron All Sky Survey, which is a joint 
project of the University of Massachusetts and the Infrared Processing and 
Analysis Center/California Institute of Technology, funded by the National 
Aeronautics and Space Administration and the National Science Foundation. JMO 
and MTR acknowledges financial support from PPARC. We thank the referee for 
useful comments.

\bsp

\label{lastpage}


\begin{thebibliography}{99}

\bibitem[\protect\citeauthoryear{Baraffe et al.}{1998}]{baraffe98}
Baraffe I., Chabrier G., Allard F., Hauschildt P., 1998, A\&A, 337, 403

\bibitem[\protect\citeauthoryear{Barrado y Navascu\'{e}s et al.}{2003}]{barrado03}
Barrado y Navascu\'{e}s D., B\'{e}jar V.J.S., Mundt R. et al., 2003, A\&A, 404, 171

\bibitem[\protect\citeauthoryear{Barrado y Navascu\'{e}s \& Mart\'{\i}n}{2003}]{barrado03b}
Barrado y Navascu\'{e}s D., Mart\'{\i}n E.L., 2003, AJ, 126, 2997

\bibitem[\protect\citeauthoryear{B\'{e}jar et al.}{2001}]{bejar01}
B\'{e}jar V.J.S., Mart\'{\i}n E.L., Zapatero Osorio M.R. et al., 2001, ApJ, 556, 830

\bibitem[\protect\citeauthoryear{Bessell \& Brett}{1988}]{bessell88}
Bessell M.S., Brett J.M., 1988, PASP, 100, 1134

\bibitem[\protect\citeauthoryear{Brandner et al.}{2000}]{brandner00}
Brandner W., Zinnecker H., Alcal\'{a} J.M. et al., 2000, AJ, 120, 950

\bibitem[\protect\citeauthoryear{Brown, de Geus \& de Zeeuw}{Brown et al.}{1994}]{brown94}
Brown A.G.A, de Geus E.J., de Zeeuw P.T., 1994, A\&A, 289, 101

\bibitem[\protect\citeauthoryear{Burningham et al.}{2005}]{burningham05}
Burningham B., Naylor T., Littlefair S.P., Jeffries R.D., 2005, MNRAS, 363, 1389

\bibitem[\protect\citeauthoryear{Caballero}{2005}]{caballero05}
Caballero J.A., 2005, AN, 326, 1007 

\bibitem[\protect\citeauthoryear{Carpenter}{2001}]{carpenter01}
Carpenter J.M., 2001, AJ, 121, 2851

\bibitem[\protect\citeauthoryear{Carpenter et al.}{2001}]{carpenteretal01}
Carpenter J.M., Hillenbrand L.A., Skrutskie M.F., 2001, AJ, 121, 3160

\bibitem[\protect\citeauthoryear{Cutri et al.}{2003}]{cutri03}
Cutri R.M., Skrutskie M.F., Van Dyk S. et al., 2003, Explanatory Supplement to the 2MASS All Sky Data Release

\bibitem[\protect\citeauthoryear{Feigelson et al.}{2003}]{feigelson03}	
Feigelson E.D., Gaffney J.A.\,III, Garmire G., Hillenbrand L.A., Townsley L., 2003, ApJ, 584, 911

\bibitem[\protect\citeauthoryear{Franciosini, Pallavicini \& Sanz-Forcada}{Franciosini et al.}{2005}]{franciosini05}
Franciosini E., Pallavicini R., Sanz-Forcada J., 2005, A\&A, 446, 501 

\bibitem[\protect\citeauthoryear{Guenther \& Emerson}{1997}]{guenther97}
Guenther E.W., Emerson J.P., 1997, A\&A, 321, 803

\bibitem[\protect\citeauthoryear{Haisch, Lada \& Lada}{Haisch et al.}{2000}]{haisch00}
Haisch K.E., Lada E.A., Lada C.J., 2000, AJ, 120, 1396

\bibitem[\protect\citeauthoryear{Haisch, Lada \& Lada}{Haisch et al.}{2001a}]{haisch01}
Haisch K.E., Lada E.A., Lada C.J., 2001a, ApJ, 553, 153

\bibitem[\protect\citeauthoryear{Haisch, Lada \& Lada}{Haisch et al.}{2001b}]{haisch01b}
Haisch K.E., Lada E.A., Lada C.J., 2001b, AJ, 121, 2065

\bibitem[\protect\citeauthoryear{Hawarden et al.}{2001}]{hawarden01}
Hawarden T.G., Leggett S.K., Letawsky M.B. et al., 2001, MNRAS, 325, 563

\bibitem[\protect\citeauthoryear{Hillenbrand et al.}{1998}]{hillenbrand98}
Hillenbrand L.A., Strom S.E., Calvet N. et al., 1998, AJ, 116, 1816

\bibitem[\protect\citeauthoryear{Jeffries et al.}{2006}]{jeffries05}
Jeffries R.D., Evans P.A., Pye J.P., Briggs K.R., 2006, MNRAS accepted, astro-ph/0512441

\bibitem[\protect\citeauthoryear{Kenyon et al.}{2005}]{kenyon05}
Kenyon M.J., Jeffries R.D., Naylor T., Oliveira J.M., Maxted P.F.L., 2005,
MNRAS, 356, 89

\bibitem[\protect\citeauthoryear{Lada et al.}{2006}]{lada05}
Lada C.J., Muench A.A., Luhman K.L., 2006, AJ in press, astro-ph/0511638

\bibitem[\protect\citeauthoryear{Lada et al.}{2004}]{lada04}
Lada C.J., Muench A.A., Lada E.A., Alves J.F., 2004, AJ, 128, 1254

\bibitem[\protect\citeauthoryear{Lada et al.}{2000}]{lada00}
Lada C.J., Muench A.A., Haisch K.E et al., 2000, AJ, 120, 3162

\bibitem[\protect\citeauthoryear{Lecar \& Sasselov}{2003}]{lecar03}
Lecar M., Sasselov D.D., 2003, ApJ, 596, L99

\bibitem[\protect\citeauthoryear{Leggett et al.}{2003}]{leggett03}	
Leggett S.K., Hawarden T.G., Currie M.J., 2003, MNRAS, 345, 144

\bibitem[\protect\citeauthoryear{Megeath et al.}{2005}]{megeath05}
Megeath S.T., Hartmann L., Luhman K.L., Fazio G.G., 2005, ApJ, 634, 113

\bibitem[\protect\citeauthoryear{Muzerolle et al.}{2003}]{muzerolle03}
Muzerolle J., Hillenbrand L., Calvet N., Brice\~{n}o C., Hartmann L., 2003, 
ApJ, 592, 266

\bibitem[\protect\citeauthoryear{Oliveira et al.}{2002}]{oliveira02}
Oliveira J.M., Jeffries R.D., Kenyon M.J., Thompson S.A., Naylor T., 2002, A\&A, 382, 22

\bibitem[\protect\citeauthoryear{Oliveira \& van Loon}{2004}]{oliveira04a}
Oliveira J.M., van Loon J.Th., 2004, A\&A, 418, 663

\bibitem[\protect\citeauthoryear{Oliveira, Jeffries \& van Loon}{Oliveira et al.}{2004}]{oliveira04b}
Oliveira J.M., Jeffries R.D., van Loon J.Th., 2004, MNRAS, 347, 1327

\bibitem[\protect\citeauthoryear{Scholz \& Eisl\"{o}ffel}{2004}]{scholz04}
Scholz A., Eisl\"{o}ffel J., 2004, A\&A, 419, 249

\bibitem[\protect\citeauthoryear{Sherry, Walter \& Wolk}{Sherry et al.}{2004}]{sherry04}
Sherry W.H., Walter F.M., Wolk S.J., 2004, AJ, 128, 2316

\bibitem[\protect\citeauthoryear{Tokunaga, Simons \& Vacca}{Tokunaga et al.}{2002}]{tokunaga02}
Tokunaga A.T., Simons D.A., Vacca W.D., 2002, PASP, 114, 180

\bibitem[\protect\citeauthoryear{Van Loon \& Oliveira}{2003}]{loon03}
van Loon J.Th., Oliveira J.M., 2003, A\&A, 405, 33

\bibitem[\protect\citeauthoryear{Uchida et al.}{2004}]{uchida04}
Uchida K.I., Calvet N., Hartmann L. et al., 2004, ApJS, 154, 439

\bibitem[\protect\citeauthoryear{White \& Basri}{2003}]{white03}
White R.J., Basri G., 2003, ApJ, 582, 1109

\bibitem[\protect\citeauthoryear{Wolk}{1996}]{wolk96}
Wolk S.J., 1996, Ph.D. thesis, State Univ. New York at Stony Brook

\bibitem[\protect\citeauthoryear{Wood et al.}{2002}]{wood02}
Wood K., Lada C.J., Bjorkman J.E. et al., 2002, ApJ, 567, 1183\bibitem[\protect\citeauthoryear{Zapatero Osorio et al.}{2002}]{osorio02}
Zapatero Osorio M.R.Z., B\'{e}jar V.J.S., Pavlenko Y. et al., 2002, A\&A, 384, 937

\bibitem[\protect\citeauthoryear{Zapatero Osorio et al.}{2000}]{osorio00}
Zapatero Osorio M.R.Z., B\'{e}jar V.J.S., Mart\'{\i}n E.L. et al., 2000, Sci, 290, 103

\end{thebibliography}
\end{document}